\documentclass[nofootinbib, twocolumn,superscriptaddress,preprintnumbers,letterpaper,natbib,nofootinbib]{revtex4-1}
\pdfoutput = 1
\usepackage{amssymb}
\usepackage{amsmath}
\usepackage{epsfig}
\usepackage[usenames,dvipsnames]{xcolor}
\usepackage[breaklinks,colorlinks,urlcolor=blue,citecolor=blue,linkcolor=magenta]{hyperref}
\usepackage{cleveref}
\usepackage{multirow}
\usepackage{capt-of}
\usepackage{siunitx}
\usepackage{graphicx}
\usepackage{slashed}
\usepackage{tabularx}
\usepackage{multirow}
\usepackage{braket}
\usepackage{amsfonts}
\usepackage{mathtools}
\usepackage{mathrsfs}
\usepackage[compat=1.1.0]{tikz-feynman}
\usepackage{subfiles}
\usepackage{slashed}
  
\usepackage[normalem]{ulem}
\usepackage[shortlabels]{enumitem}  

\usepackage{orcidlink}
\newcommand{\orcid}[1]{\orcidlink{#1}}

\usepackage{fontawesome5}
\definecolor{color_git}{rgb}{0., 0., 0.}
\newcommand{\gitlink}{\href{https://github.com/stefanmarinus/Rad2nubb}{\textsc{g}it\textsc{h}ub {\large\color{color_git}\faGithub}}}

\definecolor{tabred}{rgb}{0.8392156862745098, 0.15294117647058825, 0.1568627450980392}
\definecolor{tabblue}{rgb}{0.12156862745098039, 0.4666666666666667, 0.7058823529411765}
\hypersetup{colorlinks=true, linkcolor={tabred}, citecolor={tabblue}, urlcolor={tabblue}} 

\newcommand{\beginsection}[1]{\textit{\textbf{#1.}}---}

\makeatletter
\begin{document}
\title{Radiative corrections to two-neutrino double-beta decay}
\author{Jordy de Vries\orcid{0000-0001-5037-5108}}
\affiliation{Institute for Theoretical Physics Amsterdam and Delta Institute for Theoretical Physics, University
of Amsterdam, Science Park 904, 1098 XH Amsterdam, The Netherlands}
\affiliation{Nikhef, Theory Group, Science Park 105, 1098 XG, Amsterdam, The Netherlands}
\author{Emanuele Mereghetti\orcid{0000-0002-8623-5796}}
\affiliation{Theoretical Division, Los Alamos National Laboratory, Los Alamos, NM 87545, USA}
\author{Saad el Morabit\orcid{0009-0000-0193-8891}}
\affiliation{Institute for Theoretical Physics Amsterdam and Delta Institute for Theoretical Physics, University
of Amsterdam, Science Park 904, 1098 XH Amsterdam, The Netherlands}
\affiliation{Nikhef, Theory Group, Science Park 105, 1098 XG, Amsterdam, The Netherlands}
\author{Stefan Sandner\orcid{0000-0002-1802-9018}}
\affiliation{Theoretical Division, Los Alamos National Laboratory, Los Alamos, NM 87545, USA}

\begin{abstract} 
\noindent 
We use heavy‑nucleus effective field theory to compute radiative corrections to two-neutrino double‑$\beta$ decay ($2\nu\beta\beta$). Our main result is the first derivation of a universal radiative‑correction factor for double‑weak decays -- the analogue of the Sirlin function in single-$\beta$ decay -- independent of nuclear matrix elements and excitation energies. This “double‑weak Sirlin function” depends on the individual electron energies as well as their relative angle and differs significantly from the approximation obtained by summing two single-$\beta$ decay Sirlin functions. In addition, we calculate the nuclear-structure-dependent component of the radiative corrections and find that they can still be  neglected at current experimental sensitivities. On the other hand, the double‑weak Sirlin function induces distortions of the electron energies and angular spectra that are comparable in size to the leading nuclear‑structure correction parametrized by the ratio of nuclear matrix elements, $\xi_{31}$. Our results indicate that extractions of nuclear‑structure information and tests of the Standard Model from high‑precision $2\nu\beta\beta$ measurements must include  double‑weak radiative corrections, implying that recent extractions of $\xi_{31}$ should be revisited. 
\end{abstract}

\preprint{LA-UR-26-22239}

\maketitle

\setlength\parskip{4pt}

\beginsection{Introduction} The search for neutrinoless double‑$\beta$ decay ($0\nu\beta\beta$) is a central goal of contemporary neutrino physics. Its observation would establish lepton‑number violation and the Majorana nature of neutrinos, with far‑reaching consequences for understanding the mechanism of neutrino‑mass generation and the matter-antimatter asymmetry of our Universe~\cite{Bilenky:2014uka,Pas:2015eia,Agostini:2022zub,Adams:2022jwx}. In all current and next‑generation experiments, the Standard Model (SM) allowed two‑neutrino mode ($2\nu\beta\beta$) is both an irreducible background and, increasingly, a precision observable in its own right. It was first observed in $^{82}\mathrm{Se}$~\cite{Elliott:1987kp}, and now $2\nu\beta\beta$ half‑lives of ${}^{76}\mathrm{Ge}$, ${}^{100}\mathrm{Mo}$, ${}^{130}\mathrm{Te}$ and ${}^{136}\mathrm{Xe}$ are known at the (sub‑)percent level~\cite{GERDA:2023wbr,NEMO-3:2019gwo,CUPID-Mo:2023lru,CUORE:2025xue,KamLAND-Zen:2019imh,EXO-200:2013xfn}.

Next‑generation tonne‑scale setups aim to probe $0\nu\beta\beta$ half‑lives at the level of $10^{28}\,\mathrm{yr}$~\cite{SuperNEMO:2010wnd,LEGEND:2021bnm,nEXO:2017nam,CUPID:2019imh,NEXT:2020amj,Adams:2022jwx, AMoRE:2015asn,XLZD:2024pdv}, while recording millions of $2\nu\beta\beta$ events. This will enable high‑statistics measurements of electron energy spectra, angular correlations and other differential distributions. Sub‑percent  experimental and theoretical control of the $2\nu\beta\beta$ spectral shape is crucial for two reasons: $(i)$ it constrains nuclear‑structure calculations entering $0\nu\beta\beta$ nuclear matrix elements (NMEs)~\cite{Simkovic:2018rdz,KamLAND-Zen:2019imh,CUORE:2025xue,CUPID-Mo:2023lru}, $(ii)$ it modifies the end of the $2\nu\beta\beta$ spectrum which forms a background for $0\nu\beta\beta$ searches and $(iii)$ it opens a precision frontier for tests of the SM and searches for physics beyond the SM~\cite{Deppisch:2020mxv,Bolton:2020ncv,Bossio:2023wpj}. A recent CUORE analysis~\cite{CUORE:2025xue}, for instance, indicates tension between extracted NME ratios and theoretical predictions, underscoring the need for a consistent treatment of all relevant corrections.

On the theory side, the $2\nu\beta\beta$ spectrum is usually organized as an expansion in lepton energies over typical excitation energies of the intermediate nucleus, achieving a factorization between nuclear matrix elements and leptonic phase space~\cite{Simkovic:2018rdz}. Ref.~\cite{Morabit:2024sms} showed that subleading nuclear effects, i.e. weak magnetism and pion‑exchange double‑weak currents, induce distortions in the $2\nu\beta\beta$ spectrum that are numerically important and not accounted for in experimental analyses. In addition, a  description at the (sub-)percent level must include electromagnetic radiative corrections, which can be even larger. 

\begin{figure*}
\vspace{-0.5cm}
\centering
\includegraphics[width=\textwidth]{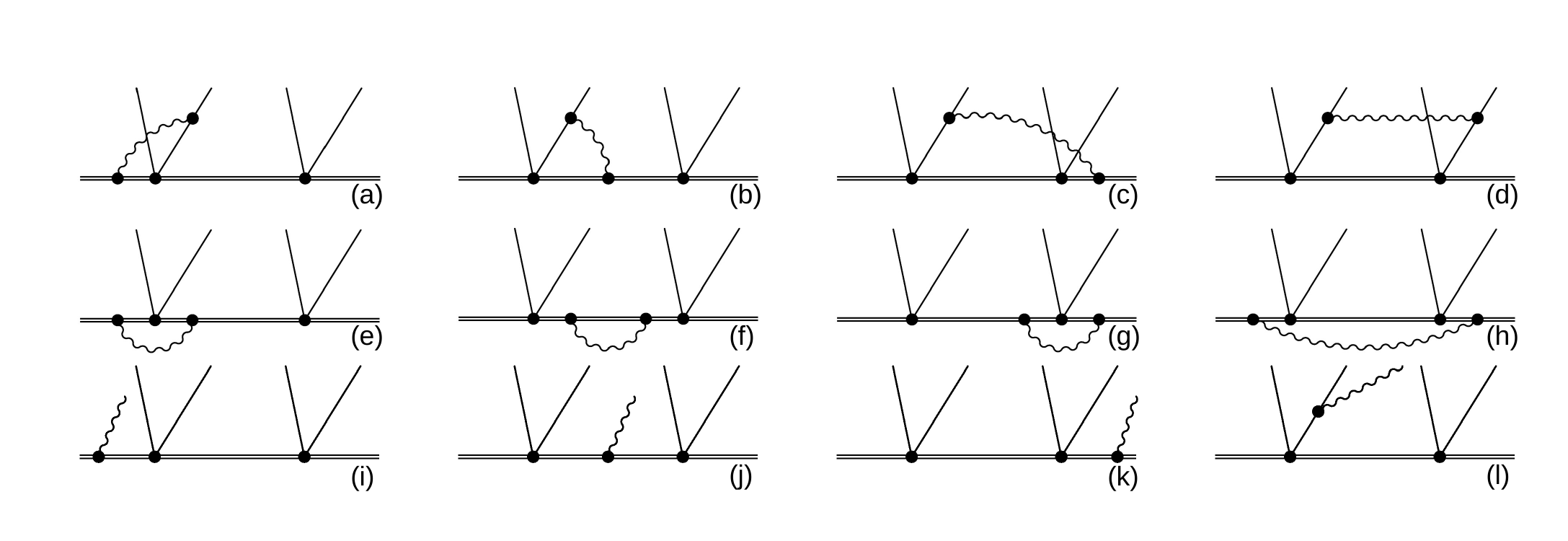}
\vspace{-0.8cm}
\caption{Diagrams contributing to $2\nu\beta\beta$ at $\mathcal O(\alpha)$. Double lines denote nuclear states in the heavy-particle EFT. Plain and wiggly lines denote leptons and photons, respectively. Black dots are vertices from the leading Lagrangian. Diagrams analogous to $(a)$, $(b)$, $(c)$, and $(l)$, but with emission from the second electron, are not shown. All possible lepton permutations are implied.}
\label{fig:virtual}
\end{figure*}

Radiative effects arise from photons with a broad range of virtualities, from hard modes that renormalize single‑nucleon couplings to long‑wavelength (ultrasoft) photons that are sensitive to global nuclear properties and the detailed kinematics of the emitted electrons. Ultrasoft radiative corrections to the spectral shape have so far only been estimated by analogy with single-$\beta$ decay~\cite{Nitescu:2024tvj}. In this Letter we provide the first explicit calculation of ultrasoft radiative corrections to double‑weak processes. Working in a heavy‑nucleus effective field theory (EFT) that includes fields for the initial and final nuclei as well as for each relevant intermediate nuclear excitation, we evaluate the full set of virtual and real diagrams at $\mathcal O(\alpha)$, see Fig.~\ref{fig:virtual}, including topologies with photons attached to the intermediate nucleus and between the two electrons that have no analogue in single-$\beta$ decay. Expanding in the ratio of the lepton and intermediate state energies, we derive a universal radiative‑correction factor for double‑weak decays -- the analogue of the Sirlin function in single-$\beta$ decay~\cite{Sirlin:1967zza}, but now depending on both electron energies and their relative angle. In addition, we compute non-universal corrections that depend on the intermediate‑state spectrum.

Our main findings are: $(i)$ the “double‑weak Sirlin function” significantly differs in most of the phase space from the approximation obtained by summing two single‑$\beta$ Sirlin functions, $(ii)$ the resulting radiative distortions of the $2\nu\beta\beta$ electron energy spectrum and angular distributions are comparable in size to the leading nuclear‑structure effect governed by $\xi_{31}$, and $(iii)$ at leading order in a large‑$\omega_n$ expansion, where $\omega_n$ denotes the nuclear excitation energy, the radiative corrections can be expressed in a compact form that is accurate at the sub‑per‑mille level for phenomenologically relevant nuclei and does not require additional nuclear structure input. These results imply that precision extractions of nuclear‑structure parameters and SM tests from $2\nu\beta\beta$ data must consistently include double‑weak radiative corrections, and they motivate dedicated searches for radiative $2\nu\beta\beta+\gamma$ events, whose branching ratios we quantify. While bremsstrahlung is included in most simulations for experimental analyses, they are modeled from scattering with the detector material but do not include radiation originating from the decay itself. This omission can now be remedied.

\beginsection{Background} We start by reviewing the theoretical description of the $2\nu\beta\beta$ spectrum, which we describe in terms of the sum and difference of the electron energies as well as their relative angle
\begin{align}
\nonumber
\epsilon = E_{e_1} + E_{e_2} - 2 m_e\,, \; \Delta = \frac{E_{e_1} - E_{e_2}}{2}\,, \; y_{12} = \hat p_{e_1} \cdot \hat p_{e_2}\,,\;
\end{align}
with $\hat p = \vec p/|\vec p\,|$. Differently from allowed $\beta$-spectra, at leading order in the multipole expansion there is no complete factorization between NMEs and leptonic physics. The NMEs depend on the lepton energies through
\begin{align}
\label{eq:LOME}
M^{K,L}_{GT} = m_e \sum_n G_n \frac{\omega_n}{\omega_n^2 - \epsilon^2_{K,L}}\,,
\end{align}
where $n$ denotes the set of $1^+$ states of the intermediate nucleus, $\omega_n = E_n - (E_f + E_i)/2$, with $E_{i}$, $E_f$, and $E_n$ the energies of the initial, final, and intermediate nuclei, and $G_n$ are the Gamow-Teller matrix elements
\begin{align}
\label{eq:GnLO}
G_n = \langle f| \sum_k  \vec \sigma_k\tau_k^+ |n\rangle \cdot \langle n | \sum_l \vec \sigma_l \tau_l^+ | i\rangle\,,
\end{align}
where $| i \rangle$, $| n \rangle$, $ | f \rangle$ refer to initial, intermediate and final states respectively. The combinations of lepton energies  $\epsilon_{K,L}$ are defined as $2\epsilon_K = E_{e_1} + E_{\nu_1} - E_{e_2} - E_{\nu_2}$, $2\epsilon_L = E_{e_1} + E_{\nu_2} - E_{e_2} - E_{\nu_1}$. Ref.~\cite{Simkovic:2018rdz} noticed that the expansion of Eq.~\eqref{eq:GnLO} in powers of $\epsilon_{K,L}/\omega_n$ converges rapidly. After expanding the energy denominators in Eq.~\eqref{eq:LOME}, the triple differential rate at leading order in the chiral expansion can be expressed as
\begin{align}
\label{totddecay}
\begin{split}
\frac{\mathrm{d}\Gamma}{\mathrm{d}\epsilon\, \mathrm{d} \Delta\, \mathrm{d} y_{12}} &= \frac{1}{2} \left(\frac{g_A}{g_V}\right)^4 \left(M_{GT}^{(-1)} \right)^2   \\
&\times \left[\frac{\mathrm{d} G^{2\nu}_0}{\mathrm{d}\epsilon\, \mathrm{d}\Delta\, \mathrm{d} y_{12}  }+ \frac{\mathrm{d} G^{2\nu}_2}{\mathrm{d}\epsilon\, \mathrm{d}\Delta\, \mathrm{d} y_{12} }\xi_{31}   + \ldots \right]\,,
\end{split}
\end{align}
where $\ldots$ denote higher-order terms. Eq.~\eqref{totddecay} achieves the factorization of the NMEs and the leptonic phase space. Here $g_A/g_V = 1.2753(13)$  denotes the nucleon axial coupling extracted from neutron decay~\cite{ParticleDataGroup:2024cfk}. With this choice, radiative corrections to $g_A/g_V$ are automatically included in the experimental value of the neutron axial coupling, but the decay rate contains factors of the nucleon vector coupling $g_V$, which equals 1 at leading order, but receives corrections at $\mathcal O(\alpha)$. The NMEs appear through
\begin{align}
M_{GT}^{(-2m-1)} =m_e (2m_e)^{2m} \sum_n \frac{G_n}{\omega_n^{2m+1}}\,,
\end{align}
and $\xi_{31} = M_{GT}^{(-3)}/M_{GT}^{(-1)}$. Information about the leptonic phase space is contained in $G^{}_{0,2}$, given in the End Matter, which are only functions of the lepton energies and relative angle. 

\beginsection{Methods} Electromagnetic corrections to $2\nu\beta\beta$ can arise from photons with different virtualities~\cite{Cirigliano:2024msg}, including high-energy photons that will affect the single-nucleon vector and axial couplings~\cite{Seng:2018qru,Seng:2018yzq,Gorchtein:2021fce,Cirigliano:2022hob,Cirigliano:2023fnz,Cirigliano:2024nfi,Tomalak:2026wks}, potential modes sensitive to nuclear structure \cite{Cirigliano:2024msg,Cirigliano:2024rfk}, and ultrasoft photons sensitive to global features of nuclei, such as their charges and radii~\cite{Ando:2004rk, Cirigliano:2022hob,Cirigliano:2024msg,Plestid:2024eib,Plestid:2025idc,Hill:2023acw,Hill:2023bfh,VanderGriend:2025mdc,Cao:2025zxs,Crosas:2025xyv}. Corrections that will affect the spectral shape are induced either by ultrasoft modes or, at higher order in an expansion in $E_e R$, where $R$ denotes scales close to the nuclear radius, by potential modes~\cite{Cirigliano:2024msg}. The latter are traditionally included via finite-size modifications of the Fermi function~\cite{Kotila:2012zza,Simkovic:2018rdz,Hayen:2017pwg}. The former have been estimated by assuming the correction to be the same as for single-$\beta$ decays~\cite{Nitescu:2024tvj}, and captured by the sum of two Sirlin functions $g(E_e,E_0)$~\cite{Sirlin:1967zza}, one for each outgoing electron, where $E_e$ and $E_0$ are the electron energy and endpoint energy.  This  cannot be exact, since it ignores the spectrum of nuclear excitations and correlations between the two electrons. Furthermore, the prescription of Ref. \cite{Nitescu:2024tvj} implicitly evaluates the Sirlin function at a fixed renormalization scale $\mu = m_p$, where $m_p$ is the proton mass, which induces a large logarithm in $g(E_e,E_0)$ and a scale dependence that is not compensated by that of the ``inner radiative corrections'' contained in $g_V(\mu)$~\cite{Czarnecki:2004cw,Cirigliano:2023fnz}. 

We compute the ultrasoft contributions to double-weak decays explicitly, by calculating loop corrections in a low-energy EFT containing as degrees of freedom fields for the initial and final state nuclei, as in single-$\beta$ decay~\cite{Plestid:2024eib,Crosas:2025xyv}, and, in addition, one field for each excitation of the intermediate nucleus. This theory is formally obtained by integrating out nucleon degrees of freedom and it is organized in an expansion in $E_e/k_F$, where we take the Fermi momentum $k_F \sim 100\,\mathrm{MeV}$  as the scale at which nucleon degrees of freedom get resolved. At leading order in this expansion, the kinetic and weak decay components of the Lagrangian assume the form~\cite{Georgi:1990um,Jenkins:1990jv,Wise:1992hn}
\begin{align}
\label{eq:HEFT}
\begin{split}
&\mathcal L_{\mathrm{EFT}} = \bar {\cal A}_i i D_0 {\cal A}_i  + \bar {\cal A}_f \left[ i  D_0 - \Delta_{fi} \right]  {\cal A}_f  \\ 
&+ \sum_n \bar{\mathcal B}^j_n  \left[i  D_0 - \Delta_{ni}\right] \mathcal B^j_n  \\ 
& - \frac{2 G_F}{\sqrt{2}} \sum_n V_{ud} \left\{ g_{A} ^{i n}\, \bar{\mathcal  B}^j_n \mathcal A_i  + g_{A}^{ fn} \bar{\mathcal A}_f \mathcal B^j_n \right\} \bar e \gamma^j P_L\nu\,,
\end{split}
\end{align}
where $\mathcal A_{i,f}$ are scalar fields for the $0^+$ initial and final states, while $\mathcal B^j_n$ denotes vector fields for the $1^+$ intermediate states and the covariant derivatives, $D$, contain minimal couplings to photons. In Eq.~\eqref{eq:HEFT} we chose the mass of the initial state as the reference mass. The mass splittings are given by $\Delta_{fi} = E_f - E_i$ and $\Delta_{ni} = E_{n} - E_i$. The nuclear axial couplings $g_{A}^{in}$ and $g_{A}^{fn}$ depend on the single nucleon axial coupling and on NMEs of the nucleon axial current. At leading order in chiral EFT 
\begin{align}
g^{in}_{A} g^{fn}_{A} = \frac{1}{3} \left(\frac{g_A}{g_V}\right)^2 G_n\,.
\end{align}
The nuclear charge and axial radii, magnetic and weak magnetic moments enter the EFT expansion at higher order in $E_e/k_F$. The Lagrangian~\eqref{eq:HEFT} can be used to calculate ultrasoft radiative corrections, where the dependence on the lepton kinematic and on the excited states energies appears explicitly in the EFT Feynman diagrams, while the dependence on nuclear structure is encoded in the couplings of the theory. While the Sirlin function in single-$\beta$ decay depends only on $E_e$ and on the $\mathcal Q$ value, the virtual and real emission diagrams contributing to $2\nu\beta\beta$ and shown in Fig.~\ref{fig:virtual} are complicated by the appearance of two additional scales: the  nuclear excitation energy, $\omega_n$, and the two-electron invariant mass, $s = (p_{e_1} + p_{e_2})^2$, which can be traded for the angle $y_{12}$. 

The calculation of the diagrams shown in Fig.~\ref{fig:virtual} is rather lengthy. Topologies like diagram $(b)$, in which the photon is absorbed by the intermediate nucleus, or diagram $(d)$, with the photon exchange between the two electrons, do not have corresponding diagrams in single-$\beta$ decays, and thus are not captured by the Sirlin function. We evaluated the diagrams for generic $\omega_n$, generalizing tensor reduction techniques~\cite{Denner:2002ii} to the heavy particle EFT integrals with linear propagators encountered in the problem. The evaluation of the virtual diagrams then reduces to the calculation of bubble and triangle relativistic integrals, and bubble, triangle, and box heavy-particle integrals. The relativistic integrals can be found in Ref.~\cite{Ellis:2007qk}. A prescription for scalar heavy particle integrals is given in Ref.~\cite{Zupan:2002je}, but we found the general expressions to be not immediately useful for our specific problem. As heavy particle triangle and box integrals are ultraviolet (UV) finite, we showed that they can be obtained by taking the heavy-particle limit of relativistic box and triangles implemented in \texttt{PackageX}~\cite{Patel:2016fam,Patel:2015tea} which can be later implemented in \texttt{FeynCalc}~\cite{Shtabovenko:2023FeynCalc,Shtabovenko:2020FeynCalc,Shtabovenko:2016FeynCalc,Mertig:1991FeynCalc}. More details will be presented in Refs.~\cite{Long,Saad}. We checked explicitly that the numerical evaluation of heavy particle integrals agrees with the limit of the relativistic integrals.  The virtual diagrams contain UV and infrared (IR) divergences which we regulate in dimensional regularization.  The UV divergences are canceled by the renormalization of $g_V$~\cite{Cirigliano:2022hob,Cirigliano:2023fnz,Cirigliano:2024nfi}, while the IR divergences explicitly cancel against those appearing in the real emission diagrams. The real emission diagrams contribute a finite part that we evaluated using subtraction techniques~\cite{Frixione:1995ms,Alioli:2010xd}. 

In addition to working with generic $\omega_n$, we obtained an expansion in the limit $\omega_n \gg \epsilon_{K,L}$ in two ways. In the first approach, we took the large-$\omega_n$ limit of the full expression, while in the second we integrated out intermediate nuclear states and matched onto an even lower energy EFT with only initial and final state nuclei, a strategy introduced in Ref.~\cite{Dekens:2025skl}. The two expressions agree at fixed order, providing an important check of our result. At leading order in $1/\omega_n$, radiative corrections modify Eq.~\eqref{totddecay} by shifting
\begin{align}
\label{eq:dGalpha}
\frac{\mathrm{d} G^{2\nu}_0}{\mathrm{d}\epsilon\, \mathrm{d}\Delta\, \mathrm{d} y_{12}  }  \mapsto  g_V^4 \frac{\mathrm{d} G^{2\nu}_0}{\mathrm{d} \epsilon\, \mathrm{d}\Delta\, \mathrm{d} y_{12}  } + \frac{\alpha}{2\pi} \frac{\mathrm{d} G^{2\nu}_\alpha}{\mathrm{d} \epsilon\, \mathrm{d} \Delta\, \mathrm{d} y_{12}  }\,, \end{align}
where $\mathrm{d} G_\alpha^{2\nu}$ contains contributions of ultrasoft modes, while hard photon modes cause the vector coupling $g_V$ to deviate from 1. The two terms on the right-hand side of Eq.~\eqref{eq:dGalpha} both depend on the renormalization scale $\mu$, in such a way that the full $\mathcal O(\alpha)$ corrections is scale independent~\cite{Cirigliano:2023fnz,Cirigliano:2024nfi}. The renormalization scale dependence of $\mathrm{d} G^{2\nu}_\alpha$ also drops out when we look at normalized spectra.  

At lowest order in $1/\omega_n$ all dependence on NMEs is absorbed in $M^{(-1)}_{GT}$, but this is no longer true at higher orders. Expanding the radiative corrections at $\mathcal O(\omega_n^{-3})$ we find not only a correction to the function $G^{2\nu}_2$, which multiplies $\xi_{31}$, but also a dependence on $\ln \mu/\omega_n$, giving rise to new NMEs different from $M_{GT}^{(-2m-1)}$. The expression containing the full $\omega_n$-dependence is too large to depict, but we give the large-$\omega_n$ expression for $\mathrm{d} G^{2\nu}_\alpha$ in the End Matter. We have checked that even for the lowest intermediate states appearing in ${}^{76}\mathrm{Ge}$ and ${}^{136}\mathrm{Xe}$, this limit agrees with the full expression within $10\%$. This is sufficient for a  description of the spectrum accurate at the $\mathcal O(10^{-3})$ level. We also provide the expression, denoted by $\mathrm{d} \hat{G}^{2\nu}_{\alpha}$, from Ref.~\cite{Nitescu:2024tvj}, which assumes the radiative corrections to be the same as in single-$\beta$ decay. 
\begin{figure}[!t]
\vspace{-0.4cm}
\includegraphics[width=0.49\textwidth]{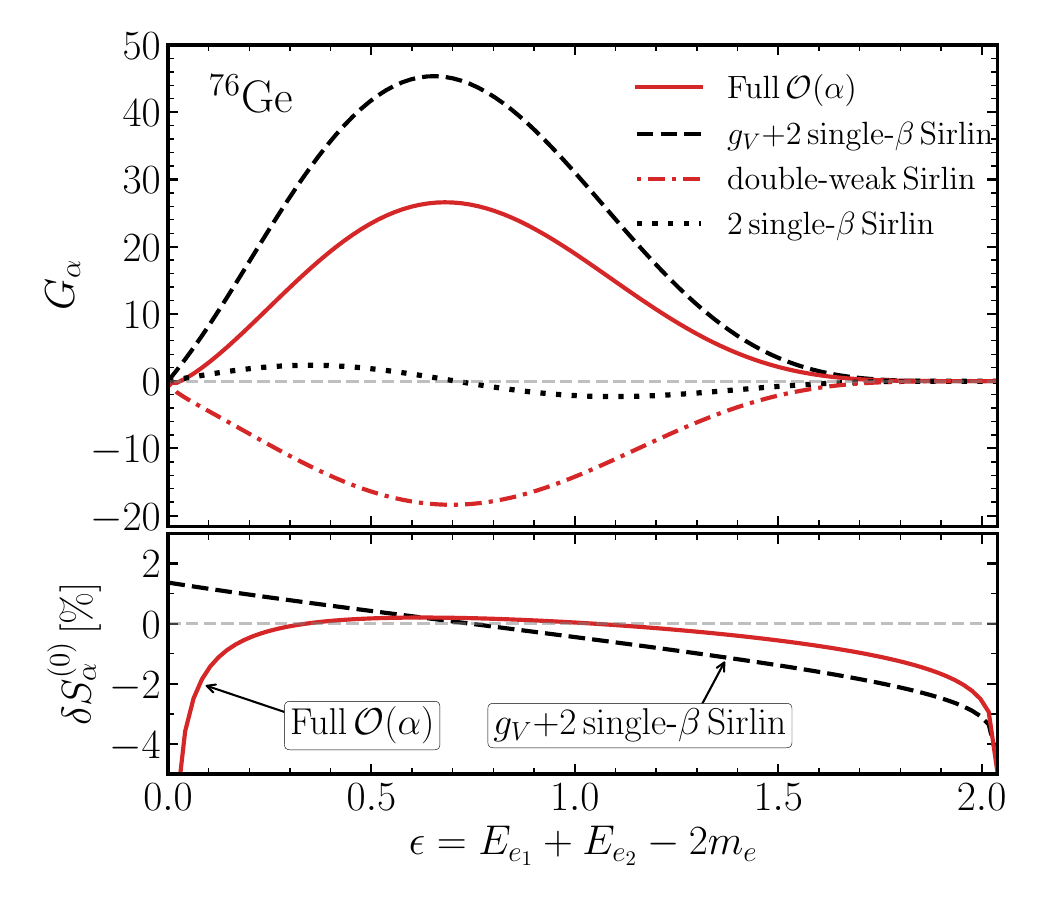}
\vspace{-0.5cm}
\caption{Top: Full calculation (solid red) and single-$\beta$ decay approximation (dashed black) of the $\mathcal O(\alpha)$ corrections to $2\nu\beta\beta$ as defined in Eq.~\eqref{eq:Ga}. We highlight the contribution from the double-weak Sirlin function (dashed-dotted red) against a naive summation of two single-$\beta$ Sirlin functions (dotted black), for fixed $\mu = 2 E_0$, see main text. Bottom: Resulting spectral distortion of the $\mathcal{O}(\alpha)$ corrections, see Eq.~\eqref{eq:deltaS}.We only show $^{76}\mathrm{Ge}$, but the behavior is similar for other isotopes.}
\label{fig:Sirlin}
\end{figure}

\beginsection{Differential Decay Rate} We start by comparing our results to the approximation of Ref.~\cite{Nitescu:2024tvj}. We define the dimensionless functions $G_\alpha$ 
\begin{align}
\label{eq:Ga}
\frac{m_e^{10}}{\ln 2}\frac{(G_F V_{ud})^4}{8 \pi^7 m_e^2}    {G}_\alpha(\epsilon) \equiv  (g_V^4 - 1)\frac{\mathrm{d} G^{2\nu}_0}{\mathrm{d}\epsilon} + \frac{\alpha}{2\pi}\frac{\mathrm{d} G^{2\nu}_\alpha}{\mathrm{d}\epsilon}\,,
\end{align}
and analogously for $\hat G_\alpha(\epsilon)$. $G_\alpha$ contains all $\mathcal O(\alpha)$ corrections and is renormalization scale independent. The differential rates $\mathrm{d} G_\alpha^{2\nu}$ and $d G_0^{2\nu}$ are shown in the End Matter, while we take $g_V$ from Refs.~\cite{Cirigliano:2023fnz,VincenzoCiriglianoPrivate}. Fig.~\ref{fig:Sirlin} compares the dimensionless radiative function, $G_\alpha(\epsilon)$, obtained from our full calculation to the approximation, $\hat G_\alpha(\epsilon)$, based on the sum of two single-$\beta$ decay Sirlin functions. The full double‑weak radiative correction differs significantly across the entire energy range. Considering the individual terms in Eq.~\eqref{eq:Ga}, corrections to $g_V$ are sizable. For example at the scale $\mu_{\mathrm{low}} = 2 E_0$, where the two-electron endpoint energy $E_0 = \mathcal Q + 2 m_e$ is chosen in analogy to the single-$\beta$ decay literature~\cite{Czarnecki:2004cw}, $g_V(\mu_{\mathrm{low}}) = 1.02054(12)$ for $^{76}\mathrm{Ge}$. This correction is important for the total (differential) rate, but does not contribute to the spectral shape. The correction coming from $\mathrm{d} G_\alpha^{2\nu}$ is energy dependent and, for $\mu = 2 E_0$,  negative over the full range in $\epsilon$. On the other hand,  $\mathrm{d} \hat G_\alpha^{2\nu}$ changes sign, and, for this choice of $\mu$, is smaller in absolute value than  $\mathrm{d} G_\alpha^{2\nu}$. As a result, using two single-$\beta$ decay Sirlin functions significantly overestimates the total radiative corrections. 

To quantify the impact on observables, we consider the normalized electron and angular spectra
\begin{align}
S^{(n)}_{i}(\epsilon)\equiv \frac{1}{\Gamma}\frac{\mathrm{d}\Gamma}{\mathrm{d}\epsilon}\,, \qquad S^{(n)}_i(y_{12})\equiv \frac{1}{\Gamma}\frac{\mathrm{d} \Gamma}{\mathrm{d} y_{12}} \,,
\end{align}
where $(n)$ labels the order in the lepton‑energy expansion and $i = \{0,\alpha \}$ denotes respectively the exclusion or inclusion of radiative corrections. These quantities are not affected by the large theoretical uncertainties on $M^{(-1)}_{GT}$ and do not depend on other normalization factors as $g_A$ and $g_V$. We characterize distortions relative to the leading‑order spectrum via
\begin{align}
\label{eq:deltaS}
\delta S^{(n)}_i(\epsilon) &= \frac{S^{(n)}_{i}(\epsilon) - S^{(0)}_0(\epsilon)}{S^{(0)}_0(\epsilon)}\,,
\end{align}
and analogously for $y_{12}$. In the bottom panel of Fig.~\ref{fig:Sirlin} we show the impact of using the full radiative correction versus two single-$\beta$ decay Sirlin functions on $\delta S_\alpha$. We see that the two corrections are of similar size, but different shape.
\begin{figure*}[!t]
\hspace{-0.5cm}
\includegraphics[width=0.99\textwidth]{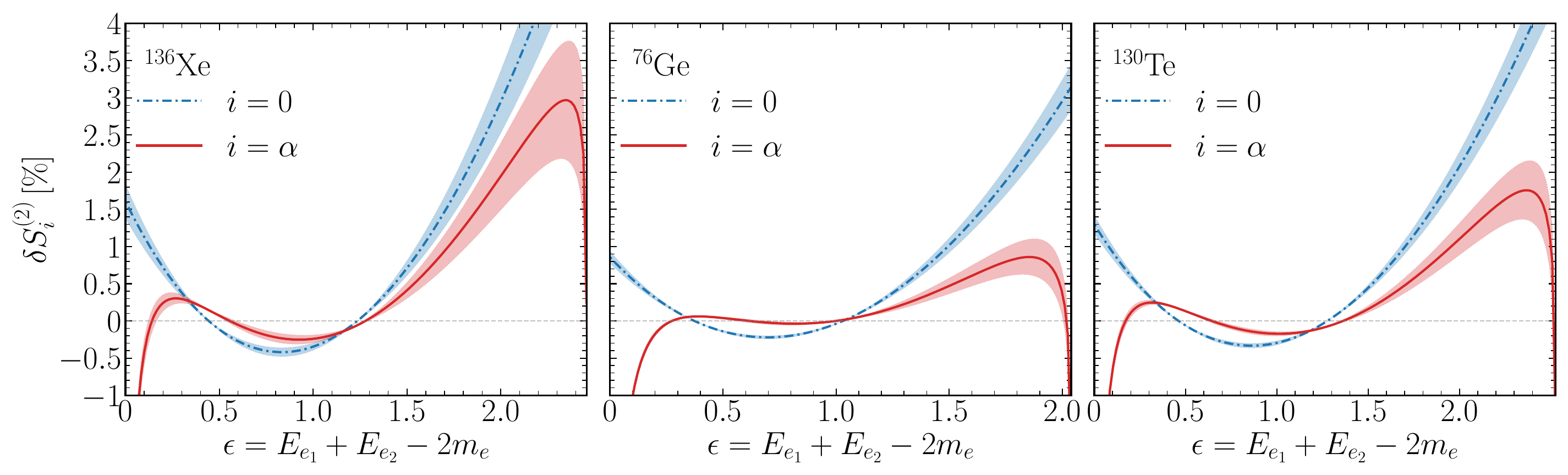}
\vspace{-0.25cm}
\caption{Interplay between radiative corrections and $\xi_{31}$ in the NSM for the electron energy distribution. The red solid line is the result of including $\mathcal{O}(\alpha)$ corrections, while the blue dashed line excludes them. Bands correspond to a $\pm10\%$ variation of $\xi_{31}$. We omit $^{100}\mathrm{Mo}$ since the corresponding $\xi_{31}$ is so large that $\mathcal{O}(\alpha)$ correction do not lead to any appreciable change.}
\label{fig:eps}
\end{figure*}
A major goal of $2\nu\beta\beta$ experiments is to extract $\xi_{31}$ to test nuclear structure methods~\cite{Simkovic:2018rdz,KamLAND-Zen:2019imh,CUPID-Mo:2023lru, CUORE:2025xue}. We therefore compare the radiative correction to the distortion induced by $\xi_{31}$, for which we use the theoretical values obtained in the Nuclear Shell Model (NSM)~\cite{Dekens:2024hlz,Castillo:2025wyr,Coraggio:2018tuo,Coraggio:2022vgy, LuigiCoraggioPrivate,CUPID-Mo:2023lru,CUORE:2025xue} and in the Quasiparticle Random Phase Approximation (QRPA)~\cite{Simkovic:2013qiy,Simkovic:2018rdz,CUPID-Mo:2023lru,CUORE:2025xue}
\begin{align}
\{ {}^{76}\mathrm{Ge}, {}^{100}\mathrm{Mo}, {}^{136}\mathrm{Xe}, {}^{130}\mathrm{Te} \} &\overset{\mathrm{\,NSM\,}}{\simeq}  \{ 0.12, 0.33, 0.12, 0.16\}  \,, \nonumber \\
\{ {}^{76}\mathrm{Ge}, {}^{100}\mathrm{Mo}, {}^{136}\mathrm{Xe}, {}^{130}\mathrm{Te} \} &\overset{\mathrm{QRPA}}{\simeq}  \{ 0.11, 0.49, 0.20, 0.32\}  \,. \nonumber 
\end{align}
As our goal is only to compare the sizes of radiative and nuclear‑structure distortions, we do not attempt a detailed error analysis of $\xi_{31}$ itself, but, for illustration purposes, in Fig.~\ref{fig:eps} we assume $\xi_{31}$ to have a $\pm10\%$ error, which roughly corresponds to the effect of varying the $g_A^{\mathrm{eff}}$ parameter in the QRPA calculation of Ref.~\cite{Simkovic:2018rdz}. Fig.~\ref{fig:eps} shows the interplay between radiative corrections and the nuclear‑structure effect using $\xi_{31}$ from the NSM. In ${}^{76}\mathrm{Ge}$, the double‑weak radiative correction almost cancels the $\xi_{31}$-induced distortion over the full energy range. For the other isotopes the cancellation remains substantial but is less complete near the endpoint. The larger values of $\xi_{31}$ in QRPA somewhat lessen the impact of the radiative distortions. A similar pattern is observed in the angular distribution in Fig.~\ref{ang} where for ${}^{100}\mathrm{Mo}$ the two effects nearly cancel. 
\begin{figure}[!t]
\includegraphics[width=0.45\textwidth]{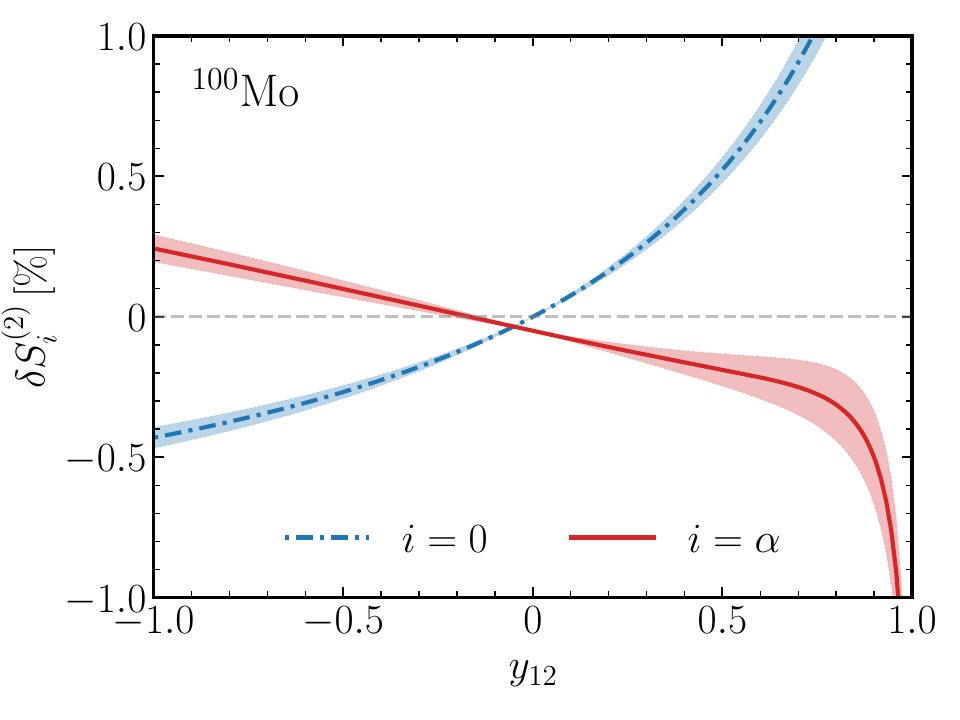}
\vspace{-0.3cm}
\caption{Interplay between radiative corrections and $\xi_{31}$ in the NSM for the angular distribution in $^{100}\mathrm{Mo}$. Color coding as in Fig.~\ref{fig:eps}. }
\label{ang}
\end{figure}
These features imply that analyses extracting $\xi_{31}$ from precision $2\nu\beta\beta$ data must include double‑weak radiative corrections. Otherwise, the omitted radiative terms can bias $\xi_{31}$ toward smaller effective values, as perhaps suggested by recent CUORE results~\cite{CUORE:2025xue}.

\begin{figure}[!t]
\centering
\includegraphics[width=0.45\textwidth]{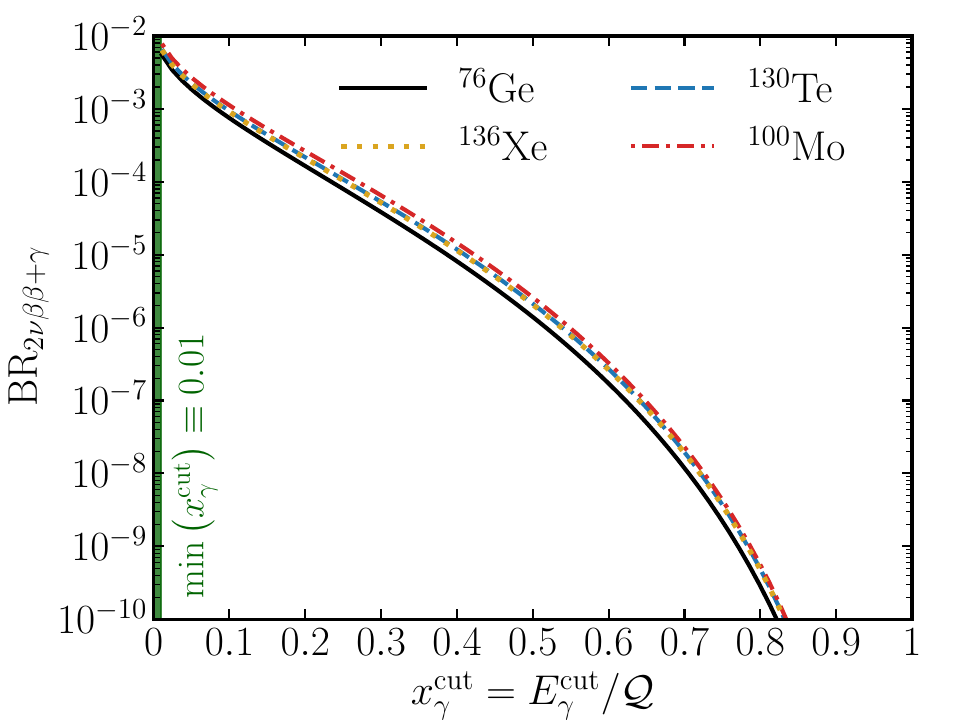}
\vspace{-0.3cm}
\caption{Branching ratio of $2\nu\beta\beta + \gamma$ as a function of the photon energy cut $x_\gamma^{\mathrm{cut}}= E_\gamma^{\mathrm{cut}}/\mathcal Q$. }
\label{fig:brdiff}
\end{figure}

\beginsection{Radiative double-weak decay} Our calculation of the real‑emission diagrams also yields the rate for the radiative process $2\nu\beta\beta$+$\gamma$. We define the branching ratio as a function of a cut on the photon energy as
\begin{align}
\mathrm{BR}_{2\nu\beta\beta + \gamma} = \frac{1}{\Gamma(2\nu\beta\beta)}\int^1_{x_\gamma^{\mathrm{cut}}} \mathrm{d} x_\gamma    \frac{\mathrm{d} \Gamma(2\nu\beta\beta + \gamma)}{\mathrm{d} x_\gamma}\,, 
\end{align}
with $x^{(\mathrm{cut})}_\gamma=E^{(\mathrm{cut})}_\gamma/\mathcal{Q}$. At leading order in the large-$\omega_n$ expansion, the branching ratio does not depend on NMEs and the dependence on $x_\gamma^{\mathrm{cut}}$ for different isotopes has a very similar behavior, see Fig.~\ref{fig:brdiff}. The exact value of $x_\gamma^{\mathrm{cut}}$ depends on the actual experimental setup and we estimate the minimal achievable value to be $\mathrm{min}\left( x_\gamma^{\mathrm{cut}}\right) \sim 0.01$. The branching ratio is $\mathcal O(10^{-2}$-$10^{-3})$ for $x_\gamma^{\mathrm{cut}} = (0.01$-$0.1)$ and falls rapidly for harder cuts, suggesting that this mode could be observable in next-generation high-statistics experiments, but will not produce a significant number of high-energy photons. 

\beginsection{Conclusion} We have carried out the first calculation of ultrasoft radiative corrections to double‑weak processes, formulating the analogue of the Sirlin function for $2\nu\beta\beta$ within a heavy‑nucleus EFT that includes intermediate nuclear excitations. The resulting “double‑weak Sirlin function” produces $\mathcal{O}(\alpha)$ corrections to $2\nu\beta\beta$ spectra that significantly differ from the approximation obtained by summing two single‑$\beta$ Sirlin functions and induces distortions comparable in size to the leading nuclear‑structure effect proportional to $\xi_{31}$. These distortions can partially cancel or mimic $\xi_{31}$ over wide regions of phase space, implying that existing and future extractions of $\xi_{31}$ and related NME ratios must include double‑weak radiative effects. Our prediction of the radiative $2\nu\beta\beta$+$\gamma$ branching ratios motivates dedicated searches for this mode. The same EFT formalism adopted here can be used to systematically include other important corrections, such as $\mathcal O(\alpha^2 Z)$ corrections and higher-order corrections in an expansion in the lepton energy over $k_F$. Looking ahead, incorporating the full double‑weak Sirlin function into experimental analyses of differential $2\nu\beta\beta$ data -- treating radiative corrections and $\xi_{31}$ on the same footing -- will sharpen nuclear‑structure constraints relevant for $0\nu\beta\beta$ and turn precision $2\nu\beta\beta$ measurements into a sensitive probe of electroweak dynamics in nuclei. 

\beginsection{Code availability} We release a python jupyter notebook containing all results presented in this work. It can be found here:~\gitlink.

\beginsection{Acknowledgments} We thank Auke-Pieter Colijn, Patrick Decowski, Maxime Pierre and Michael Graesser for useful discussions. We thank Luigi Coraggio and Javier Men\'endez for clarifications on the shell model nuclear matrix elements. We thank  Ryan Bouabid, Wouter Dekens, Javier Men\'endez, Carmen Romo-Luque and Ralph Massarczyk for comments on the manuscript. We acknowledge support from the DOE Topical Collaboration “Nuclear Theory for New Physics” award No. DE-SC0023663. E.M. and S.S. are supported by the U.S. Department of Energy Office and by the Laboratory Directed Research and Development (LDRD) program of Los Alamos National Laboratory under project numbers 20230047DR,  20250164ER and 20260246ER. Los Alamos National Laboratory is operated by Triad National Security, LLC, for the National Nuclear Security Administration of the U.S. Department of Energy (Contract No. 89233218CNA000001)
\bibliography{biblio}
\begin{widetext}
\cleardoublepage
\section*{End Matter}
We present the expressions for the triple differential $2\nu\beta\beta$ decay rates at tree level and at $\mathcal O(\alpha)$. At tree level, the lepton energy dependence is described by the functions 
\begin{align}
\label{dGamma}
\frac{\mathrm{d} G^{2\nu}_{n}}{\mathrm{d}\epsilon\, \mathrm{d}\Delta\, \mathrm{d} y_{12} } = \frac{1}{\ln 2}\frac{(G_F V_{ud})^4}{8 \pi^7 m_e^2}  \int^{\mathcal{Q}-\epsilon}_{0} \mathrm{d} E_{\nu_1}  E^2_{e_1}  E^2_{e_2}  E_{\nu_1}^2 E_{\nu_2}^2  \beta_1 \beta_2 F(E_{e_1},Z_f) F(E_{e_2},Z_f)  \mathcal A_n\,,
\end{align}
where $\vec{\beta}_{1(2)} \equiv \vec{p}_{e_{1(2)}}/E_{e_{1(2)}}$, $\beta_{1(2)} \equiv |\vec{\beta}_{1(2)}|$ and energy conservation fixes $E_{\nu_2} = \mathcal Q - \epsilon - E_{\nu_1}$. $F$ is the Fermi function~\cite{Fermi:1934hr}, which includes the leading Coulomb corrections. We use here the expression given in Refs.~\cite{Simkovic:2018rdz,Hayen:2017pwg}.
 
The Fermi functions depends on the nuclear radius $R$, for which we choose $R = \sqrt{\frac{5}{3} \langle r^2 \rangle }$, with the charge radii taken from  Ref. \cite{Angeli:2013epw}. While the total rate depends on the choice of $R$, the spectral distortion is largely insensitive. The functions $\mathcal A_n$ are given by
\begin{align}
\label{eq:A02}
\mathcal A_0  = 1 - \beta_1 \beta_2 y_{12} \,, \qquad 
\mathcal A_2  = (1 - \beta_1 \beta_2 y_{12} ) \frac{\epsilon_K^2 + \epsilon_L^2}{(2m_e)^2} \,.
\end{align}
The integral over $E_{\nu_1}$ can be carried out explicitly, yielding 
\begin{align}
\label{dGamma}
\frac{\mathrm{d} G^{2\nu}_{0}}{\mathrm{d}\epsilon\, \mathrm{d}\Delta\, \mathrm{d} y_{12} } = \frac{1}{\ln 2}\frac{(G_F V_{ud})^4}{8 \pi^7 m_e^2}  \frac{1}{30}  E^2_{e_1}  E^2_{e_2}    \beta_1 \beta_2 (E_0 - E_{e_1} - E_{e_2})^5  F(E_{e_1},Z_f)  F(E_{e_2},Z_f) (1 - \beta_1 \beta_2 y_{12})\,,
\end{align}
where $E_0 = \mathcal{Q} + 2m_e$. Similarly the $n=2$ expression can be integrated analytically over $E_{\nu_1}$.

We now turn to the expression for the radiative correction in the large-$\omega_n$ limit. After having performed the integration over the neutrino energies, and for the real emission phase space, over the photon energy, we obtain
\begin{align}
\label{eq:largew}
&\frac{\mathrm{d} G^{2\nu}_\alpha}{\mathrm{d}\epsilon\, \mathrm{d}\Delta\, \mathrm{d} y_{12}  } = \frac{1}{\ln 2}\frac{(G_F V_{ud})^4}{8 \pi^7 m_e^2}  \frac{1}{30}  E_{e_1}^2 E_{e_2}^2 \beta_1 \beta_2 \,  (E_0 - E_{e_1} - E_{e_2})^5  F(E_{e1},Z_f)  F(E_{e2},Z_f) \nonumber \\ 
\times & \Bigg\{ (1-\beta_1 \beta_2 y_{12}) \left[3 L_\mu- 8 f(\beta_1) - 8 f(\beta_2)  + 3 L(\beta_1)  + 3 L(\beta_2) +  2 \frac{s-2 m_e^2}{s} f_3(\beta_{12}) +  I_\varepsilon(p_{e_1}, p_{e_2}) \right.\nonumber \\
& \left. +  \left( 6  - 2 L(\beta_1) - 2 L(\beta_2)+ 2 \left(1 - \frac{2 m_e^2}{s}\right) L(\beta_{12})\right)   \left( \ln \frac{ m_e^2}{4(E_0 - E_{e_1} - E_{e_2})^2} + \frac{137}{30} \right)\right] \nonumber \\
& -  \frac{2 m_e^2}{E_{e_1} E_{e_2}} \beta_{12}^2 L(\beta_{12}) - \frac{2 m_e^2}{E_1^2}L(\beta_1) - \frac{2 m_e^2}{E_2^2} L(\beta_{2}) \nonumber \\
& + \frac{(E_0 - E_{e_1} - E_{e_2})}{6} \Bigg[\frac{L(\beta_1) }{E_{e_1}}  \left(2 + \frac{m_e^2}{E_{e_1} E_{e_2}} - \left(2+  \frac{m_e^2}{E^2_{e_1}}\right) \frac{\beta_2}{\beta_1}y_{12} \right) \nonumber \\
 &+\frac{L(\beta_2)}{E_{e_2}} \left(2 + \frac{m_e^2}{E_{e_1} E_{e_2}} - \left(2+  \frac{m_e^2}{E^2_{e_2}}\right) \frac{\beta_1}{\beta_2}y_{12} \right) - 6 \left(\frac{1}{E_{e_1}} + \frac{1}{E_{e_2}} - y_{12} \left(\frac{\beta_1}{E_{e_2} \beta_2} +\frac{\beta_2}{E_{e_1} \beta_1}\right) \right) \Bigg] \nonumber \\
&+ \frac{(E_0 - E_{e_1} - E_{e_2})^2}{42} \Bigg[\frac{\beta_1 -  y_{12} \beta_2}{E^2_{e_1} \beta_1}   L(\beta_1) +\frac{ \beta_2 -  y_{12} \beta_1}{ E^2_{e_2} \beta_2}    L(\beta_2)  - \frac{4}{E_{e_1} E_{e_2}} + \frac{2y_{12} }{\beta_1 \beta_2}  \left( \frac{1}{E_{e_1}^2} +\frac{1}{E^2_{e_2}}  - \frac{2 m_e^2}{E_{e_1}^2 E_{e_2}^2}\right)\Bigg]\Bigg\}\,.
\end{align}
In this expression $L_\mu = \ln \mu^2/m_e^2$. $f(\beta)$ and $L(\beta)$ are functions that also appear in the Sirlin function
\begin{align}
f(\beta) =  \frac{1}{\beta} {\rm Li}_2\left(\frac{2\beta}{1+\beta}\right) + \frac{1}{4 \beta} \ln^2 \frac{1+\beta}{1-\beta}\,, \qquad    L(\beta) = \frac{1}{\beta} \ln \frac{1+\beta}{1-\beta}\,,
\end{align}
with $\mathrm{Li}_2$ the dilogarithm, while $f_3$ arises from a relativistic triangle diagram with two massive lines~\cite{Ellis:2007qk}
\begin{align}
f_3(\beta_{12}) &=\frac{1}{\beta_{12}} \Bigg[- 2 \mathrm{Li}_2(-x_{12}) -2 \ln (-x_{12})\, \ln (1+x_{12})  + \frac{1}{2} \ln^2 ( -x_{12}) - \frac{2\pi^2}{3}\Bigg]\,,
\end{align}
with $\beta_{12} = (1 - 4 m_e^2/s)^{1/2}$ and $x_{12}=(\beta_{12} -1)/(\beta_{12} +1)$. The function $I_\varepsilon$ arises from the electron's real emissions and can be reconstructed from Ref.~\cite{Alioli:2010xd}. We define the auxiliary quantities
\begin{align}
a &= \beta_1^2 + \beta_2^2 - 2 \vec\beta_1 \cdot \vec\beta_2\,, &\;    b &= \frac{\beta_1^2 \beta_2^2 - (\vec\beta_1 \cdot \vec \beta_2)^2}{a}\,, &\;  c &= \sqrt{\frac{b}{4a}}\,, &\; x_1 &= \frac{\beta_1^2 - \vec\beta_1 \cdot \vec\beta_2}{a}\,, \nonumber\\ 
x_2 &= \frac{\beta_2^2 - \vec\beta_1 \cdot \vec \beta_2}{a}\,, &\; z_1 &= \frac{\sqrt{x_1^2 + 4 c^2} - x_1}{2c}\,, &\; z_2 &= \frac{\sqrt{x_2^2 + 4 c^2} + x_2}{2c}\,, &\;    z_\pm &= \frac{1\pm \sqrt{1-b}}{\sqrt{b}}\,, \nonumber
\end{align}
the function $I_\varepsilon$ can be expressed as
\begin{align}
I_\varepsilon = \left(K(z_2) - K(z_1) \right) \frac{1 - \vec \beta_1 \cdot \vec \beta_2}{\sqrt{a (1-b)}}\,,
\end{align}
and
\begin{align}
K(z) &=  -\frac{1}{2} \ln^2 \frac{(z - z_-)(z_+ - z)}{(z_+ + z)(z_- + z)}  - 2 {\rm Li}_2\left(\frac{2 z_- (z_+ - z)}{(z_+ - z_-) (z_- + z)}\right) - 2 {\rm Li}_2\left(-\frac{2 z_+ (z_- + z)}{(z_+ - z_-) (z_+ - z)}\right).
\end{align}

In the approximation that radiative corrections are captured by the sum of two single-$\beta$ decay Sirlin function, Ref.~\cite{Nitescu:2024tvj} obtained
\begin{align}
\label{eq:sirlin0}
\frac{\mathrm{d} \hat{G}^{2\nu}_\alpha}{\mathrm{d}\epsilon\, \mathrm{d}\Delta} =   \frac{\mathrm{d} {G}^{2\nu}_0}{\mathrm{d}\epsilon\, \mathrm{d}\Delta} \left( g(E_{e_1}, E_0 - m_e) + g(E_{e_2}, E_0 - E_{e_1}) \right)\,,
\end{align}
with no information on the angular dependence. For consistency with our heavy particle approach, we will use here the Sirlin function as computed in Heavy Baryon Chiral Perturbation Theory~\cite{Cirigliano:2022hob}
\begin{align}
g(E_e, E_0) =   \frac{3}{2} L_\mu  - 4 f(\beta) + 2 \ln \frac{m_e^2}{4\bar E^2} + 8 - \frac{4}{3} \frac{\bar E}{E_e}  +  L(\beta)  \left( - \ln \frac{m_e^2}{4 \bar E^2} -2 + \beta^2 + \frac{\bar E^2}{12 E_e^2} + \frac{2}{3} \frac{\bar E}{E_e}\right),
\end{align}
with $\bar E = E_0  - E_e$.
This differs from the Sirlin function in Ref.~\cite{Sirlin:1967zza} by a constant and by the fact that $\ln m_p$ is replaced by the logarithm of the renormalization scale $\mu$.

The final ingredient needed for the calculation of the radiative corrections is the value of $g_V$ at different renormalization scales. We follow here the extraction of Ref.~\cite{Cirigliano:2023fnz}, which provides the value at $\mu = m_p$ and the evolution kernel. A few representative values are reported in Tab.~\ref{tab:gv}.
\begin{table}[!t]
\begin{tabular}{|c|c|c|c|c|c|c|c|c|c|}
\hline
$\mu$ ($\mathrm{MeV}$) & $m_e$ & 6.12 & 7.0 & 8.1 & 25  & 50 & 250 & 500 & $m_p$\\ 
\hline
$g_V -1$ (\%) & 2.499 &2.054 & 2.031 & 2.003 & 1.802  & 1.678 & 1.390 & 1.266 & 1.153\\\hline
\end{tabular}
\caption{Representative values of the vector coupling $g_V$ as a function of the renormalization scale. The second, third and fourth scale correspond to the choice $\mu = 2 E_0$, using the $\mathcal Q$ values of $^{76}$Ge, $^{136}\mathrm{Xe}$ and $^{100}\mathrm{Mo}$, respectively.}
\label{tab:gv}
\end{table}

Finally, the decay rate for the radiative process $2\nu\beta\beta+\gamma$ is given by
\begin{align}
\frac{\mathrm{d} \Gamma(2\nu\beta\beta+\gamma)}{\mathrm{d} E_\gamma} &= \frac{1}{2} g_A^4 \left(M_{GT}^{(-1)} \right)^2  \frac{1}{\ln 2}\frac{(G_F V_{ud})^4}{8 \pi^7 m_e^2} \,  \frac{\alpha}{2\pi} \\
&\times\int_0^{\mathcal Q - E_\gamma} \mathrm{d} \epsilon \int^{\epsilon/2}_{-\epsilon/2} \mathrm{d} \Delta \int_{-1}^1 \mathrm{d} y_{12}    \, \frac{1}{30} E^2_{e_1}  E^2_{e_2}    \beta_1 \beta_2  (E_0 - E_{e_1} - E_{e_2} - E_\gamma)^5  F(E_{e_1},Z_f) F(E_{e_2},Z_f)\mathcal A_\gamma\,, \nonumber
\end{align}
with the amplitude 
\begin{align}
\mathcal A_\gamma &=\frac{4}{E_\gamma} \Bigg\{ -3 (1 - \beta_1 \beta_2 y_{12} ) - \frac{3}{2} \frac{E_\gamma}{E_{e_1}} \left(1- y_{12} \frac{\beta_2}{\beta_1}\right) - \frac{3}{2} \frac{E_\gamma}{E_{e_2}} \left(1- y_{12} \frac{\beta_1}{\beta_2}\right) - \frac{E_\gamma^2}{E_{e_1} E_{e_2}} \left(1 - \frac{y_{12}}{2} \left(\frac{E_{e_1} \beta_1}{E_{e_2}\beta_2} + \frac{E_{e_2} \beta_2}{E_{e_1}\beta_1} \right)\right) \nonumber \\
& + \left[ 1 - \beta_1 \beta_2 y_{12} + \frac{E_\gamma}{2 E_{e_1}} \left(1 + \frac{m_e^2}{2 E_{e_1} E_{e_2}} - y_{12}\frac{\beta_2}{\beta_1} \left(1+\frac{m_e^2}{2 E_{e_1}^2} \right)\right) + \frac{E_\gamma^2}{4 E_{e_1}^2} \left(1 - y_{12} \frac{\beta_2}{\beta_1}\right) \right] L(\beta_1) \nonumber \\ 
& + \left[ 1 - \beta_1 \beta_2 y_{12} + \frac{E_\gamma}{2 E_{e_2}} \left(1 + \frac{m_e^2}{2 E_{e_1} E_{e_2}} - y_{12}\frac{\beta_1}{\beta_2} \left(1+\frac{m_e^2}{2 E_{e_2}^2} \right)\right) + \frac{E_\gamma^2}{4 E_{e_2}^2} \left(1 - y_{12} \frac{\beta_1}{\beta_2}\right) \right] L(\beta_2) \nonumber \\
& - \frac{2 E_{e_1} E_{e_2}}{s} (1- \beta_1 \beta_2 y_{12})^2 L(\beta_{12}) \Bigg\}\,.
\end{align}
\end{widetext}
\end{document}